\documentstyle[aps,prl,twocolumn,epsfig,floats]{revtex}

\begin{document}
\bibliographystyle{prsty}

\draft 

\title{Distributions of the Conductance \\ and its Parametric 
Derivatives in Quantum Dots} 

\author{A.  G.  Huibers, S.  R.  Patel, 
and C.  M.  Marcus} \address{Department of Physics, Stanford 
University, Stanford, California 94305} 
\author{P. W. Brouwer} \address{Department of Physics, Harvard 
University, Cambridge, MA 02138} 

\author{C.  I.  Duru\"oz and 
J.  S.  Harris, Jr.} \address{Department of Electrical Engineering, 
Stanford University, Stanford, California 94305} \date{\today} 
\maketitle

\begin{abstract}
Full distributions of conductance through quantum dots with 
single-mode leads are reported for both broken and unbroken 
time-reversal symmetry.  Distributions are nongaussian and 
agree well with random matrix theory calculations that account 
for a finite dephasing time, $\tau_{\varphi}$, once broadening due 
to finite temperature $T$ is also included.  Full 
distributions of the derivatives of conductance with respect to gate 
voltage $P(dg/dV_{g})$ are also investigated.

\end{abstract}

\pacs{72.70.+m, 73.20.Fz, 73.23.-b}

The remarkable success of random matrix theory (RMT) and other 
noninteracting theories in describing the statistics of quantum transport in 
mesoscopic electronic systems is surprising considering that 
electron-electron interactions not accounted for in the theory are 
sizable compared to other energy scales in these systems.  In the last 
few years, an essentially complete statistical theory of mesoscopic 
fluctuation and interference effects in disordered or chaotic 
(irregularly shaped) quantum dots and wires has been formulated using 
these methods \cite{Beenakker1996,Efetov1997,ChaosSol} and has recently 
been extended beyond the ``universal'' regime to include 
short-trajectory effects \cite{beyondRMT}, clarifying the degree to 
which these theories are indeed universal \cite{AltshulerSimons}.

But how seriously should a noninteracting theory be taken when 
describing real metallic or semiconductor structures?  Apparently, 
this depends on the quantity being measured.  For instance, the mean 
and variance of mesoscopic conductance fluctuations in open quantum 
dots and disordered wires appear to be in good agreement with random 
matrix theory \cite{Beenakker1996} and sigma model calculations 
\cite{Efetov1995} once 
temperature and dephasing effects are included, whereas the 
distribution of energies needed to add subsequent electrons to a 
closed dot does not appear distributed according to the famous 
Wigner-Dyson law, a basic result of RMT\cite{SSS}.

In this Letter we carry out a stringent test of statistical theories of 
mesoscopic conductance fluctuations by measuring the full distribution 
$P(g)$ of conductance, $g$, through chaotically shaped ballistic 
quantum dots.  Conductance distributions of quantum dots have 
previously been calculated within RMT
\cite{Jalabert1994,Baranger1994}, including the effects of 
dephasing \cite{Beenakker1996,Baranger1995,Brouwer1995,Brouwer1997}.  
The distributions are universal for any fully chaotic or 
disordered dot, sensitive only to whether time reversal symmetry is 
obeyed ($\beta=1$) or broken ($\beta=2$), controlled by adding a 
magnetic field, $B > \,\sim \phi_o/A_{dot}$, where 
$A_{dot}$ is the dot area and $\phi_o = h/e$ is the flux quantum.  
RMT yields quite interesting (i.e.\ strongly nongaussian) 
distributions when one or two quantum modes connect the dot to bulk 
reservoirs.  To date, however, experimental measurements of these 
nongaussian distributions have not been reported, first because it is 
difficult to generate large ensembles of statistically identical 
devices, and second because dephasing, which acts roughly as extra 
modes coupling the dot to the environment, leads to nearly gaussian 
distributions \cite{Brouwer1997,Chan1995,McCann1997}.  To see the nongaussian
distributions,  dephasing rates and temperatures comparable to the quantum level 
spacing are required.  We solve the first problem, of obtaining large 
ensembles, by using electrostatic shape distortion of gate-defined 
quantum dots in a GaAs/AlGaAs heterostructure \cite{Chan1995}.

We find good agreement between the experimental distributions and the 
RMT predictions over a broad range of temperatures once thermal 
averaging is properly accounted for.  The agreement is particularly 
surprising since we are investigating the case of single-mode leads, $N=1$.  
This is the transition between open and closed dots; for any lower 
conductance to the dot, electron-electron interactions in the form of 
Coulomb blockade dominate transport, leading to dramatic departures 
from a noninterating picture \cite{Kouwenhoven1997}.  Here $N$ 
denotes the number of modes or channels in both the left and right 
leads, giving dimensionless lead conductances $g_{\rm l} = g_{\rm r} = 
2 N$, where $g$ is in units of $e^2/h$.

\begin{figure}[bth]
\epsfxsize=\columnwidth \epsfbox{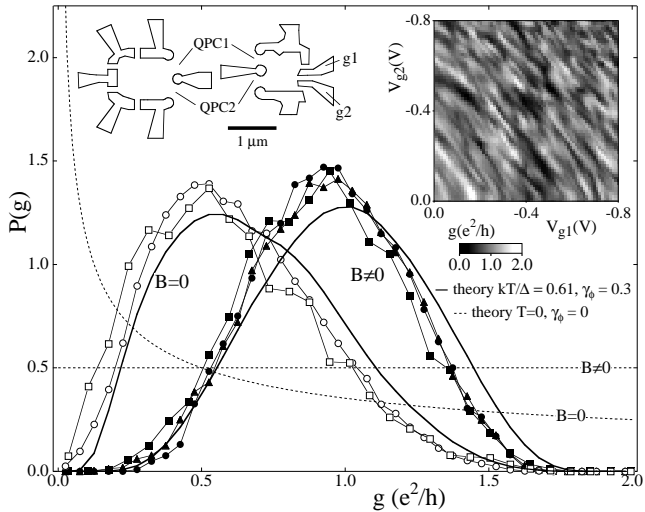} 
  \caption{ Conductance 
  distributions for $B=0$ (open circles), 40 mT (filled circles) and 60 mT 
  (filled triangles) for the $0.5\;\mu{\rm m}^2$ device at 100 mK, 
  and for $B=0$ (open squares) and 25 mT (filled squares) 
  for the $1.0\;\mu{\rm m}^2$ device at 45 mK, along with the theoretical 
  distributions for $kT/\Delta = 0.61$, $\gamma_\varphi = 0.3$
  (solid curves) and $T=0$, $\gamma_\varphi = 0$ (dashed curves).  
  Upper left inset: Pattern of gates defining each quantum 
  dot.  Upper right inset: Conductance through $0.5\;\mu{\rm m}^2$ dot 
  as a function of the two shape-distorting gates $V_{g1}$ and 
  $V_{g2}$.  }
  \label{fig1}
\end{figure}

Distributions of mesoscopic conductance at $T=0$ may be calculated 
within RMT for any number of modes $N$ in the leads using the Landauer 
formula, $g = 2 \;Tr(tt^{\dagger})$, by assuming that the $2N \times 
2N$ scattering matrix $S=\left( {\matrix{r&t'\cr t&r'\cr }} \right)$ 
is a random unitary matrix reflecting the ergodicity of chaotic 
scattering \cite{Beenakker1996}.  For single-mode leads, $N=1$, this 
calculation gives $P(g)={1 \over 2} (g/2)^{-1+\beta/2}$ 
\cite{Jalabert1994,Baranger1994}, shown as dashed lines in Fig.  
\ref{fig1}.  Notice that the $\beta=1$ distribution is skewed toward 
smaller conductance, with average conductance $\langle g 
\rangle_{\beta=1} = 2/3$, while the $\beta=2$ distribution is constant 
between $0$ and $2$ with $\langle g \rangle_{\beta=2} = 1$.  The lower 
average conductance for $\beta=1$ results from coherent 
backscattering, analogous to weak localization, at $B=0$.

Dephasing, or the loss of quantum coherence, can be modeled within RMT by
expanding the scattering matrix $S$ to include  a fictitious voltage lead that
supports a number of modes 
$\gamma_{\varphi}=2\pi \hbar / (\tau_\varphi \Delta)$, where $\Delta 
={2\pi \hbar ^2} / {m^*A_{dot}}$ is the spin-degenerate mean level 
spacing and $\tau_\varphi $ is the characteristic dephasing time 
\cite{Baranger1995,Brouwer1995}.  A recent improvement to the 
voltage-probe model that accounts for the spatially distributed nature 
of the dephasing process considers the 
limit of a voltage lead supporting an infinite number of modes, each 
with vanishing transmission, allowing a continuous value for the 
dimensionless dephasing rate $\gamma_\varphi$ \cite{Brouwer1997}.  
As described below, the general effect of dephasing is to make $P(g)$ 
narrower and roughly gaussian, and to reduce the difference in mean 
conductance upon breaking time reversal symmetry, $\delta g = \langle 
g \rangle_{\beta=2} - \langle g \rangle_{\beta=1}$.

Measurements on two lateral quantum dot with areas $0.5\;\mu{\rm m}^2$ 
($\Delta = \rm 14\;meV$) and $1.0\;\mu{\rm m}^2$ ($\Delta = \rm 
7.1\;meV$) are reported.  The devices (see Fig.\ 1 inset) are defined 
using Cr/Au depletion gates 90 nm above a two-dimensional electron gas 
(2DEG) formed at a GaAs/Al$_{0.3}$Ga$_{0.7}$As heterointerface.  
Multiple gates are used to allow independent 
control of the two point-contact leads as well as dot shape.  Sheet 
density $n=2\times 10^{11}\;\rm cm^{-2}$ and mobility $\rm \mu 
=1.4\times 10^5\;{cm^2}/{Vs}$ give an elastic mean free path of $\sim 1.5
\;\mu$m, larger than all device dimensions, so that transport within the dots
is ballistic.  The dots  were measured in a dilution refrigerator over a range
of electron  temperatures from 45 mK to 750 mK using standard 4-wire lock-in 
techniques at 43 Hz (13 Hz) and less than 7\ $\mu$V (2\ $\mu$V) bias 
voltage for the $0.5\;\mu{\rm m}^2$ ($1.0\,\mu{\rm m}^2$) device.  
Electron temperature was determined from a fit to the average Coulomb 
blockade peak width, as described in \cite{Folk1996}.

Experimental conductance distributions $P(g)$ for $| B | \ll 
\phi_{o}/A_{dot}$ ($\beta=1$) and $| B | \gg \phi_{o}/A_{dot}$ ($\beta=2$) 
are shown in Fig.\ \ref{fig1}.  Each histogram contains 
$\sim 1000$ independent samples measured at fixed $B$ drawn from the 
random landscape of conductance fluctuations in the space of 
$V_{g1}$ and $V_{g2}$ (inset Fig.\ \ref {fig1}).  Note that the 
conductance landscape appears random and the average roughly constant 
over the shape-distortion landscape.  The asymmetry of the $\beta=1$ 
distribution is striking, in contrast with previous measurements 
\cite{Chan1995,Lee1997} which found roughly gaussian distributions due 
to thermal averaging and dephasing. The average conductance at 
$\beta=2$ is $\sim e^2/h$ as expected from RMT, with a small (4\%) 
deviation at the lowest temperatures possibly due to imperfect quantum 
point contacts enchanced by incipient charging effects 
\cite{Furusaki1995,Aleiner1997}.  Once time-reveral symmetry is 
broken, the distribution becomes insensitive to magnetic field for the 
relatively small fields used in the experiment (the cyclotron 
radius is always larger than the dot size).  For instance, though $g(B)$
are uncorrelated at $B=40$ mT and 60 mT, $P(g)$ at 
these magnetic field values are nearly identical 
($\phi_{o}/A_{dot}$ is $\sim 8$ mT for the $0.5\;\mu\rm m^2$ device).

\begin{figure}[bth]
 \epsfxsize=\columnwidth \epsfbox{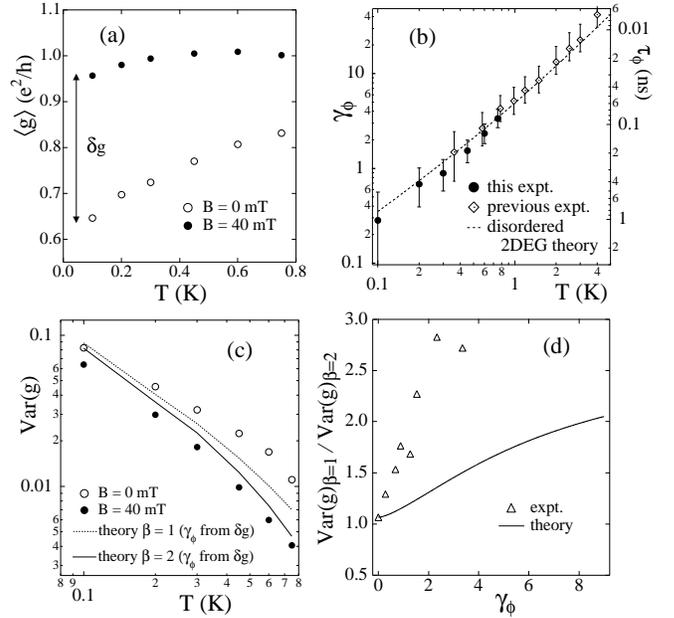} 

  \caption{ (a) Average conductance $\langle g \rangle$ as a function 
  of temperature $T$ for $B=0$ and 40 mT for the 0.5\ $\mu\rm m^2$ 
  dot.  (b) Normalized dephasing rate $\gamma_\varphi$ and dephasing 
  time $\tau_\varphi$ determined from $\delta g(T)$ for the 0.5\ 
  $\mu\rm m^2$ dot.  Note agreement with previously measured 
  $\gamma_\varphi(T)$ in a 0.4\ $\mu\rm m^2$ dot \protect\cite{Huibers1997}.
  (c) Variance of 
  conductance for $B=0$ and $B=40$ mT, corresponding to expected 
  variance for $\beta = 1$ (dashed) and $\beta = 2$ (solid) 
  including thermal smearing and dephasing effects (see text).  (d) 
  Variance ratio ${\rm Var}\:g_{\beta=1}/{\rm Var}\:g_{\beta=2}$ as a 
  function of dephasing rate $\gamma_{\varphi}$.}

  \label{fig2}
\end{figure}

Before the experimental distributions can be compared to theory it is  necessary
to measure the dephasing rate, since $P(g)$ depends on 
$\gamma_{\varphi}$.  Values for $\gamma_{\varphi}$ are measured from  the weak
localization correction to the average conductivity, $\delta  g = \langle
g\rangle _{\beta=2} - \langle g\rangle _{\beta=1}$ using  the results of Ref.\
\cite{Brouwer1997}, as shown in Fig.\ \ref{fig2}(b).  The values for
$\gamma_{\varphi}(T)$ agree with  previous measurements found using a variety of
magnetotransport  methods including weak localization and the power spectrum of 
conductance fluctuations
\cite{Huibers1997}, and extend those  measurements to lower temperatures. 
Unfortunately, while $\langle  g\rangle$ depends on temperature {\em only}
through $\gamma_{\varphi}$  (which is why weak localization is particularly
useful for measuring  dephasing), the full distributions $P(g)$ depend on
temperature both  implicitly through dephasing and explicitly through thermal
averaging.   The combined effects of dephasing and thermal smearing must in
general  be evaluated numerically, which we do as follows.  A set of values $y 
= \sum\nolimits_i {w_{i}(T)x_i}$ is generated by summing independent  samples
$x_{i}$ drawn from the known distribution $P(x, 
\gamma_{\varphi})$ \cite{Brouwer1997}, weighted by the derivative of the Fermi
function, 
$w_i = \tilde \Delta f'([i+\delta]\tilde \Delta)$, where $f'(\epsilon)  = {d
\over d \epsilon} (1+e^{\epsilon /kT})^{-1}$, $\delta$ is a  binning offset, and
$\tilde \Delta$ is the level broadening, itself  dependent on $\gamma_{\varphi}$
as described below.  By sampling over  ensembles of $x$ values, a distribution
$P(y)$ is obtained (the result  is insensitive to the choice of $\delta$ for
sufficiently large $T$).   Note that neither fluctuations in level spacing nor
fluctuations in  the coupling between the levels and modes in the leads are
included in  this simple model.

Both dephasing and temperature averaging tend to make $P(g)$ roughly 
gaussian, in which case $P(g)$ can be characterized by its mean and 
variance.  As discussed above, the mean, $\langle g \rangle$, is not 
affected by thermal averaging.  The variance is reduced both by dephasing, 
well approximated by the interpolation formula ${\rm Var}\,x = 
(a+b\gamma_{\varphi})^{-2}$ where $a =\sqrt {3}\;(\sqrt{45/16})$ and 
$b= 1\;(\sqrt{1/3})$ for $\beta =2 \;(1)$ \cite {Baranger1995}, and 
by thermal averaging, ${\rm Var}\,y = \sum\nolimits_i 
w_i^2 \:{\rm Var}\,x$.  At temperatures exceeding the level broadening this sum
can be well approximated by an integral,

\begin{equation}
{\rm Var}\,y = \tilde \Delta \Bigl[\int_{-\infty}^\infty 
[f'(\epsilon)]^2 d\epsilon \Bigr]\,{\rm Var}\,x = {\tilde \Delta \over 
{6 k T}} \,{\rm Var}\,x.
\label{varianceintegral}
\end{equation}
The integral form differs from the sum by less than 1\% for $kT 
\geq 0.6 \tilde \Delta$.  For the devices studied in this paper, 
this condition is easily satisfied, and Eq. (1) is applicable
at all measured temperatures.

The thermal averaging procedure given above takes energy intervals 
of size $\tilde \Delta$ to be statistically independent.  Dephasing 
contributes to level broadening by an amount proportional 
to the dephasing rate, which can be taken into account by defining 
the level broadening to be

\begin{equation}
	{\tilde \Delta = \Delta \left(1+\gamma_\varphi/2\right)}.
	\label{effectivedelta}
\end{equation}

\begin{figure}[bth]
  \epsfxsize=\columnwidth \epsfbox{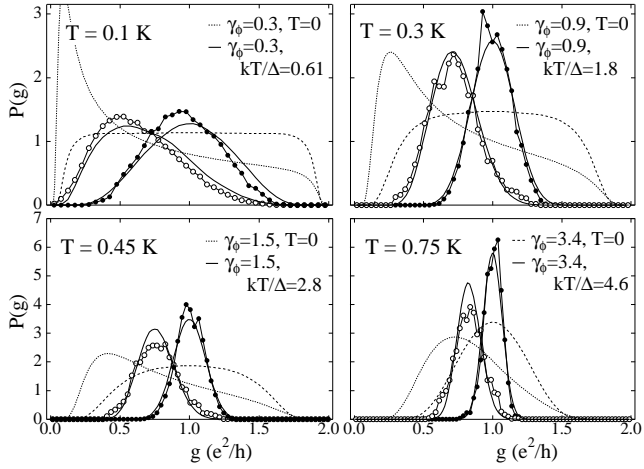}
  \caption{ Measured conductance distributions for $B=0$ (open 
  circles) and 40 mT (closed circles) for the 
  $0.5\,\mu\rm m^2$ dot at four temperatures.  Curves show 
  theoretical $\beta = 1$ (dotted) and $\beta = 2$ (dashed)
  distributions for $T=0$ and for
  measurement temperatures (solid).  $T=0$ distributions are 
  determined using the method of Ref.  \protect\cite{Brouwer1997}.  Thermal 
  distributions are calculated according to the sampling procedure 
  described in the text.  }
  \label{fig3}
\end{figure}

Inserting this definition into Eq. (1) reproduces 
a number of previously obtained results for Var $g$ in various limits: 
($\gamma_{\varphi} \ll 1$, $T \ll \Delta$) 
\cite{Jalabert1994,Baranger1994}
; ($\gamma_{\varphi} \ll 1$, $T \gg 
\Delta$)\cite{Efetov1995}; ($\gamma_{\varphi} \gg 1$, $T \ll \Delta$) 
\cite{Baranger1995,Brouwer1995}; and ($\gamma_{\varphi} \gg 1$, $T \gg 
\Delta$) \cite {Efetov1995}.  The measured variances of the 
conductance distributions for $\beta=1$ and $\beta=2$ as a function of 
temperature are compared to our thermal averaging model in Fig.\ 2(c).  
The two are in good overall agreement, however, the ratio of 
variances, ${\rm Var}\:g_{\beta=1}/{\rm Var}\:g_{\beta=2}$ shows 
significant disagreement between experiment and theory which remains 
unexplained.  In particular, the experimental ratio of variances is 
considerably larger than predicted, as seen in Fig.  \ref{fig2}(d).  
This ratio is an interesting quantity because, like $\delta g$, it 
does not suffer thermal averaging within the simple model considered 
here.
Despite the disagreement in the ratio of 
variances, the RMT results for $P(g)$ are generally in very good 
agreement with experiment across a broad range of temperatures, as 
seen in Fig. \ref{fig3}.

\begin{figure}[bth]
\epsfxsize=\columnwidth \epsfbox{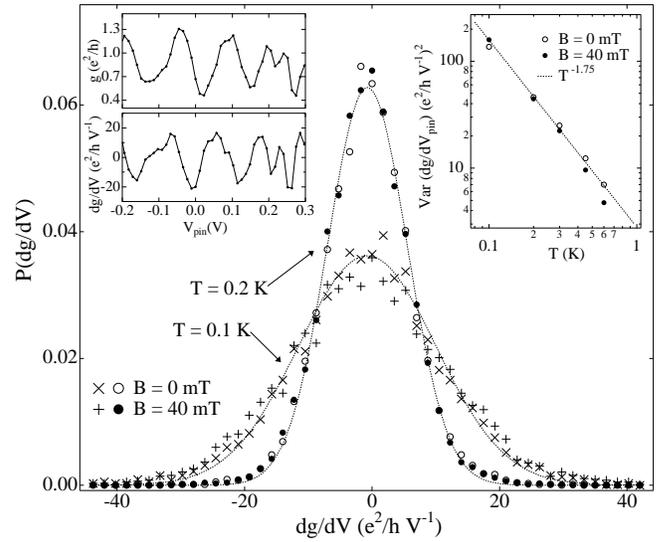} 
\caption{ Conductance 
  derivatives $P(dg/dV_{g})$ at $B = 0$ and 40 mT for $T=0.1$ K and 
  $T=0.2$ K.  Upper right inset shows Var($dg/dV_{g}$) as a function 
  of temperature, which exhibits a $T^{-1.75}$ dependence.  Upper left 
  inset shows fluctuating conductance $g$ and derivative $P(dg/dV_{g})$ 
  as function of gate voltage.  }
  \label{fig4}
\end{figure}

We have also investigated distributions of parametric derivatives of 
conductance with respect to magnetic field $dg/dB$ and gate voltage, 
$dg/dV_{g}$.  These quantities are of considerable interest as they 
are the open-system analogs of the well-studied ``level velocities'' 
$dE/dX$ of the energy levels in closed quantum chaotic systems 
\cite{AltshulerSimons}.  Distributions of parametric derivatives of 
conductance have recently been investigated theoretically using an RPA 
``charged fluid'' model, which gave interesting nongaussian distributions for both 
$\beta=1$ and $\beta=2$ \cite{Brouwer1997b}.  We have investigated 
both $P(dg/dB)$ and $P(dg/dV_{g})$ in single-mode dots, and find both 
distributions are well described by gaussians due to dephasing and thermal 
averaging.  Here we focus on $P(dg/dV_{g})$, shown in Fig.\ 4 for zero 
and nonzero magnetic field, at $T=0.1$ mK and $0.2$ mK. Thermal averaging 
dominates the width of the distribution.  We find ${\rm 
Var}\,g\propto\,T^{-1.75}$.  Distributions of $dg/dB$ are also roughly 
gaussian, with variances of 9.65 $\rm{mT}^{-2}$ at $T=0.1$ K 
and 0.939 $\rm{mT}^{-2}$ at $T=0.5$ K. To observe deviations 
from gaussian distributions of parametric derivatives, lower 
temperatures than those needed to see nongaussian $P(g)$ are required.

In summary, we have measured distribution of conductance and its 
derivatives using large ensembles of shape-deformable chaotic GaAs 
quantum dots with single-mode leads.  Accurate control of lead 
conductances and low temperature allow the nontrivial 
predictions of RMT to be observed.  We find that a thermally averaged 
RMT provides a good  description of the measured distributions, though some
features, in  particular the ratio of variances ${\rm Var}\:g_{\beta=1}/{\rm 
Var}\:g_{\beta=2}$, are inconsistent with the present model and remain to be 
resolved.
 
We thank I. Aleiner and K. Matveev for useful discussions.  We 
gratefully acknowledge support at Stanford from the Army Research 
Office under Grant DAAH04-95-1-0331, the Office of Naval Research YIP 
program under Grant N00014-94-1-0622, the NSF-NYI and PECASE programs, 
the A. P. Sloan Foundation, and support for AGH from the Fannie and 
John Hertz Foundation.  We also acknowledge support from JSEP under
Grant DAAH04-94-G-0058.  PWB acknowledges support from the NSF under
grants DMR 94-16910, DMR 96-30064, and DMR 94-17047.

\end{document}